\documentclass[twocolumn,showpacs,preprintnumbers,amsmath,amssymb]{revtex4}

\usepackage{graphicx}

\begin{document}


\title{A single photoelectron transistor
for quantum optical communications}
\author{Hideo Kosaka$^{1\ast }$, Deepak S. Rao$^{1}$, Hans D. Robinson$^{1}$,
Prabhakar Bandaru$^{1}$, Kikuo Makita$^{2}$, Eli Yablonovitch$^{1}$}
\address{$^{1}$ Electrical Engineering Department, University of California
Los Angeles, Los Angeles, CA, 90095-1594\\
$^{2}$ Photonics and Wireless Device Research Laboratories, NEC Corporation,%
\\
34 Miyukigaoka, Tsukuba, Ibaraki 305-8501, Japan\\
$^{\ast }$ On leave from Fundamental Research Laboratories, NEC Corporation.}
\date{\today}

\begin{abstract}
A single photoelectron can be trapped and its photoelectric charge detected
by a source/drain channel in a transistor. Such a transistor photodetector
can be useful for flagging the safe arrival of a photon in a quantum
repeater. The electron trap can be photo-ionized and repeatedly reset for
the arrival of successive individual photons. This single photoelectron
transistor (SPT) operating at the $\lambda$ = 1.3$\mu$m tele-communication
band, was demonstrated by using a windowed-gate double-quantum-well
InGaAs/InAlAs/InP heterostructure that was designed to provide near-zero
electron g-factor. The g-factor engineering allows selection rules that
would convert a photon's polarization to an electron spin polarization. The
safe arrival of the photo-electric charge would trigger the commencement of
the teleportation algorithm.
\end{abstract}

\maketitle

Quantum information can take several different forms and it is beneficial to
be able to convert among the different forms. One form is photon
polarization, and another is electron spin polarization.

Photons are the most convenient medium for sharing quantum information
between distant locations. Quantum key distribution \cite{Bennett92} has
been demonstrated by sending photons through a conventional optical fiber up
to distances over 80km \cite{Hiskett}. As the distance increases, the secure
data rate decreases, owing to photon loss. To expand the distance
dramatically, it is necessary to realize a quantum repeater, that is based
on quantum teleportation \cite{Bennett93}. A quantum repeater requires
quantum information storage \cite{van Enk}, and electron spin is a good
candidate for such a quantum memory. We need to have a photodetector that
converts from photon to electron, while transferring the quantum information
from photon polarization to electron spin. This is sometimes called an
entanglement preserving photodetector \cite{Vrijen}. In addition, the
photodetector must provide a trigger signal to flag the arrival of a
photo-electric charge, and to commence the teleportation algorithm.

A Field Effect Transistor (FET), and a Single Electron Transistor (SET)
based on quantum dots, can both function as sensitive electrometers that can
detect a single trapped electric charge. Our goal is to safely trap a
photoelectron, so that its spin state can then be monitored. In this paper
we demonstrate the trapping and manipulation of individual photoelectrons,
but we have not yet measured the trapped electron's spin properties.
Previous experiments have demonstrated interband photon absorption resulting
in the trapping of photo{\it holes}; on self-assembled InAs quantum dots
\cite{shields}, or on DX centers \cite{kosaka02}, near an FET source/drain
channel. These produce positive photoconductivity, that is fairly common.
The trapping of photo{\it electrons} is much more rarely observed, since it
is accompanied by negative photoconductivity \cite{rose}.

\begin{figure}
\includegraphics[width=8.1cm]{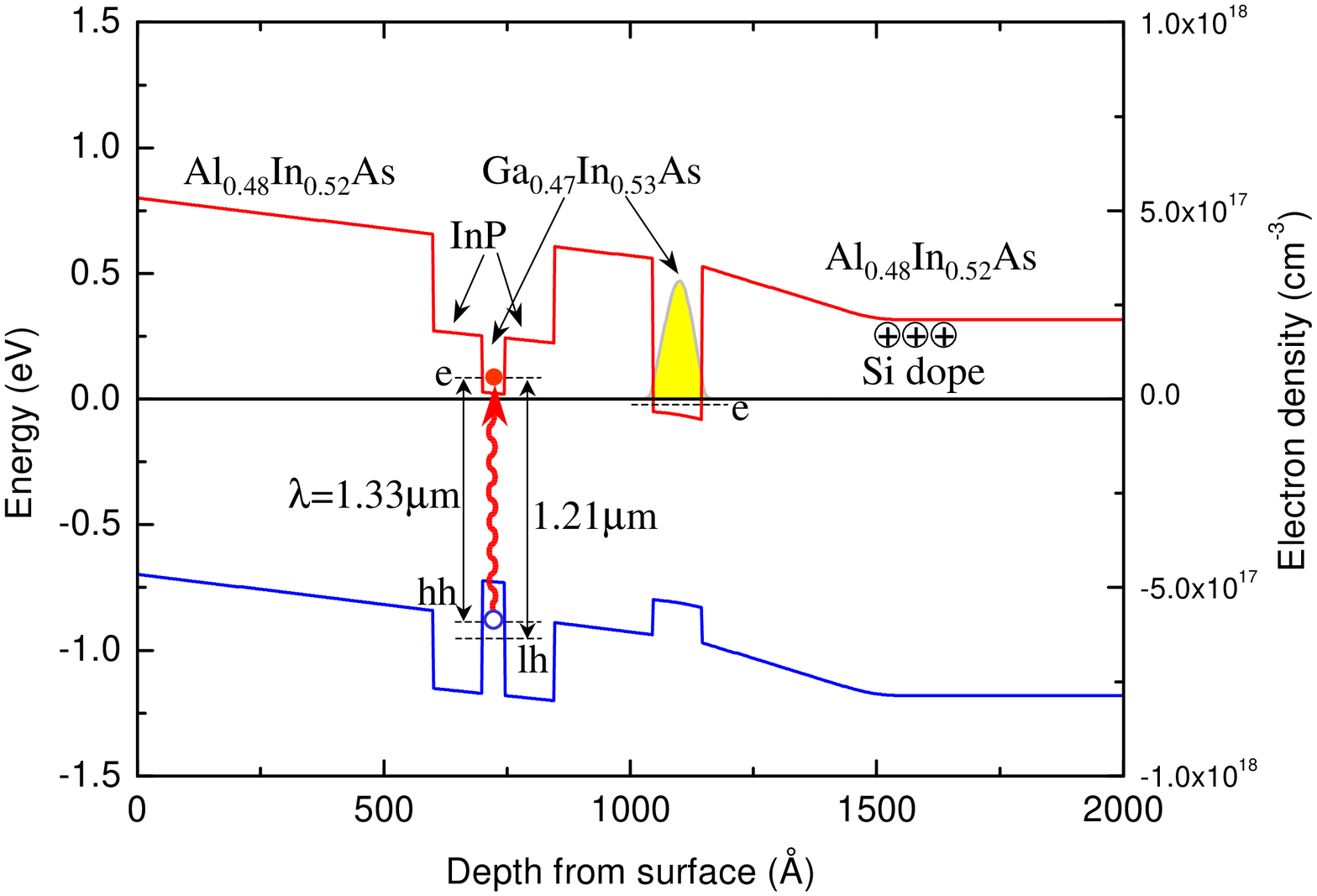}
\caption{\label{fig1} The energy band diagram of the Single Photoelectron Transistor (SPT)
at zero bias simulated by using one-dimensional Poisson/Schr\"{o}dinger
equation. Photo-induced transitions between the heavy hole band and the
conduction band is shown with an arrow. Photo-ionization of donors by $%
\lambda$ = 1.77 $\mu$m-light modulation dopes the channel. The tunneling
time of trapped electrons in the top quantum well leaking to the bottom
quantum well is estimated to be over 1{\nobreakspace}hour by WKB simulation.}
\end{figure}

Several kinds of photon effects on SETs made on modulation-doped
semiconductors have been reported. Photon assisted tunneling is the most
common effect. The tunneling takes place between an island and source-drain
reservoir \cite{blick98,blick95}, between two adjacent islands \cite{vaart},
or between an inner island and an outer ring split into Landau levels by a
magnetic field \cite{komiyama}. In all these cases, the rather long photon
wavelengths are controlled by the electron sub-band energy difference,
rather than by the fundamental bandgap as in our experiments.

These types of single photon detectors should be distinguished from
avalanche photo-diodes, where the single photon sensitivity arises from
avalanche gain. In the FET and SET photodetectors, a single trapped electric
charge can influence the current of millions of electrons in the
source/drain channel. This is indeed the mechanism of ``photoconductive
gain'' \cite{rose} that is also sometimes called ``secondary
photoconductivity'' \cite{rose}. But this form of gain can also be
considered as arising from transistor action. Thus the name ''single
photoelectron transistor'' (SPT) might be appropriate. Since the
photoelectron is safely trapped, and is known to have a long spin lifetime
in many semiconductors \cite{awschalom}, it can then be interrogated to
determine its spin state. The initial goal is to monitor the photo-electric
charge in such a way as to not disturb its spin state. Ultimately the goal
is to measure its spin state as well.

\begin{figure}
\includegraphics[width=8.1cm]{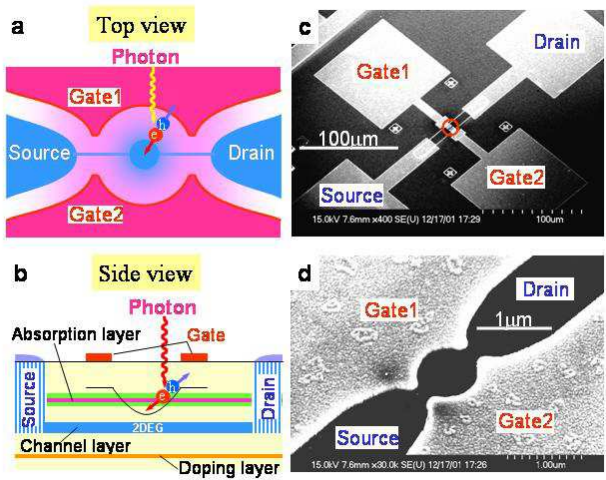}
\caption{\label{fig2} A single photoelectron transistor (SPT) with
window-gate double-quantum-well modulation-doped heterostructure.
(a), Top view of the window-gate part of the SPT. The center of
the window gates is relatively positive to the surroundings when
negative voltage is applied to the gates because of Fermi level
pinning. The blue regions indicate the two-dimensional electron
gas (2DEG) in the channel layer. (b{\bf )}, Cross-section view of
the layers in the SPT. The upper quantum well (QW) functions as an
absorption layer and lower QW serves as a 2DEG channel layer,
which is connected to source and drain. The curve on the
absorption layer illustrates the electron potential when negative
voltage is applied to the gates. (c), Scanning electron micrograph
(SEM) picture of the SPT. (d), Close-up SEM picture of the window
gate part (circled part in c). The window diameter is 1 $\mu$m.}
\end{figure}

At least three requirements should be satisfied to make a photodetector for
quantum repeaters: (1) The wavelength that should be in the 1.3$\mu $m or
1.55$\mu $m, the low-loss window of optical fibers. (2) The sign of the
photoconductivity that should be negative, which means the trapped
information carrier should be an electron instead of a hole. (3) The
electron g$_{e}$-factor, which should be small, to make the up-and-down
electron spin states as indistinguishable as possible \cite{Vrijen}. The 1st
requirement suggests interband transition rather than intraband transition.
The 2nd requirement suggests creation of a positively charged trap for an
electron. The 3rd requirement is satisfied through g$_{e}$-factor
engineering \cite{kosaka01,kiselev02}.

The single photoelectron transistor (SPT) that we present in this paper
satisfies all of the above requirements. An InGaAs quantum well is used with
a bandgap corresponding to $\lambda $ = 1.3 $\mu $m, as shown in Fig.{%
\nobreakspace}1. In Fig.{\nobreakspace}2 is shown the window-shaped circular
gates that are negatively biased above the two-dimensional electron gas
(2DEG), leaving behind a relatively positive central island. The InGaAs
absorption layer, which has a g$_{e}$-factor{\nobreakspace}={\nobreakspace}%
--4.5 in the bulk, is sandwiched between InP cladding layers, of g$_{e}$%
-factor{\nobreakspace}={\nobreakspace}+1.2, to make the effective g$_{e}$%
-factor in the absorption layer nearly zero. The measurements showed clear
evidence for negative persistent photoconductivity steps. The abrupt drops
in photoconductivity are strongly correlated with photon injection at the $%
\lambda $ = 1.3 $\mu $m wavelength, leading to the conclusion that the SPT
detects a single photon by sensing the charge of a safely trapped
photoelectron in the absorption quantum well.

The photo-absorption layer is located above the source/drain channel layer,
and both are made of In$_{0.53}$Ga$_{0.47}$As, separated by a high electron
barrier layer made of In$_{0.52}$Al$_{0.48}$As to prevent leakage. The
source/drain channel layer is modulation doped and formed into a
1-dimensional electron gas (1DEG) channel whose conductance is sensitive to
the charge state of the island in the absorption layer above it. All layers
were grown by gas-source molecular beam epitaxy on semi-insulating InP, and
consisted of a nominally undoped InP buffer layer 100nm thick; an i-In$%
_{0.52}$Al$_{0.48}$As buffer 1000nm thick; a Si-doped (5 $\times $ 10$^{17}$%
/cm$^{3}$) n-In$_{0.52}$Al$_{0.48}$As doping layer 10nm thick; an i-In$%
_{0.52}$Al$_{0.48}$As lower spacer layer 30nm thick; an i-In$_{0.53}$Ga$%
_{0.47}$As channel layer 10nm thick; an i-In$_{0.52}$Al$_{0.48}$As barrier
layer 20nm thick; an i-InP cladding layer 10nm thick; an i-In$_{0.53}$Ga$%
_{0.47}$As absorption layer 4.5nm thick; an i-InP cladding layer 10nm thick;
and an i-In$_{0.52}$Al$_{0.48}$As capping layer 60nm thick. The
modulation-doped double-quantum well structure creates a 2DEG in the lower
quantum well that is shaped into a 1DEG channel by the two split gates. The
gates surround a circular window, 1 $\mu $m in diameter, that masks out
unnecessary light exposure, and fixes the potential at the edges surrounding
of the window. The Schottky gates, Al/Pt/Au, are fabricated using
electron-beam lithography and electron-gun evaporation. The source/drain
ohmic contacts are made of AuGe/Ni/Au. Scanning Electron Microscope pictures
of the whole device and the window gates are shown in Figs.{\nobreakspace}2c{%
\nobreakspace}and{\nobreakspace}2d, respectively. The energy band diagram at
zero bias, simulated by one-dimensional Poisson/Schr\"{o}dinger equation, is
shown in Fig.{\nobreakspace}1.

The sample is illuminated by monochromatic light through a large-core glass
fiber, that is carefully shielded to block any photons from the outer
jacket. The light is created by a tungsten lamp and then filtered by a
monochromator, a long-pass filter passing wavelengths $\lambda${\tt >}%
1000nm, and a 30dB neutral density filter. The optical power at the end of
the fiber is measured by a InGaAs detector. The illumination area in the
plane of the device is about 5mm in diameter owing to light diffraction from
the end of the fiber. Given the small device active area of 7.9 $\times$ 10{%
\-}$^{9}${\nobreakspace}cm$^{2}$, defined by the 1 $\mu$m diameter gate
window, we estimate the actual light power in the active area to be 2.8 $%
\times$ 10$^{-}$$^{8}$ times smaller than the total power (assuming a
Gaussian profile). The incident photon number is estimated by multiplying
this scaling factor by the measured power divided by the photon energy.

By applying a negative voltage to the split window gates, the source/drain
current through the channel layer is pinched off. Simultaneously, the
applied negative voltage creates a two-dimensional potential minimum in the
window at the absorption layer. This is because the surface Fermi level in
the circular area is pinned by the extrinsic surface states \cite{chou}. The
electric field in the electrostatic potential well can separate an
electron-hole pair created by a photon. The electron is attracted to the
potential minimum at on center, and the hole is attracted to the negative
gates as schematically shown in Fig.{\nobreakspace}2b.

\begin{figure}
\includegraphics[width=8.1cm]{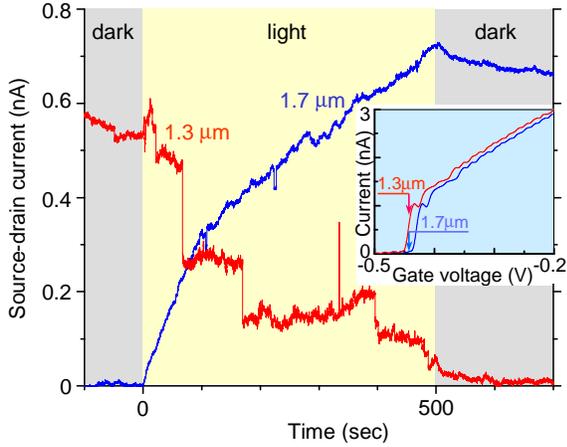}
\caption{\label{fig3} Negative persistent photoconductivity of the
SPT to $\lambda $ = 1.3 $\mu $m light starting with finite
conductance, and positive photoconductivity at $\lambda $ = 1.7
$\mu $m light starting with zero conductance. The source-drain
current drops in discrete steps when the SPT
is exposed to $\lambda $ = 1.3 $\mu $m. The inset shows the initial current{%
\nobreakspace}-{\nobreakspace}gate voltage characteristics
(I$_{sd}$-V$_{g}$
curves) and bias points for the $\lambda $ = 1.3 $\mu $m exposure and the $%
\lambda $ = 1.7 $\mu $m exposure. The $\lambda $ = 1.3 $\mu $m
photons create photoelectrons in the quantum well, which are
trapped and pinch off the 2DEG, step by step. In contrast, the
$\lambda $ = 1.7 $\mu $m photoionize the electrons and increase
the 2DEG density. Photon number absorbed in the window area is 0.3
per second, on average.}
\end{figure}

The source/drain current is measured at a constant voltage drop (V$_{sd}$)
of 0.5mV, at a temperature of 4.2{\nobreakspace}K. The interesting property
of these photodetectors is that $\lambda $ = 1.77 $\mu $m light produces
positive photoconductivity effectively doping the channel, and $\lambda $ =
1.3 $\mu $m light produces negative photoconductivity. We attribute the
channel doping by $\lambda $ = 1.77 $\mu $m light to be due to
photo-ionization of donors in the n-InAlAs doping layer. As a normal
practice, we initially prepare the photodetectors for use by means of a deep
soak in $\lambda $ = 1.77 $\mu $m light, to fully ionize the donors and to
populate the source/drain channel. The pinch-off behavior in the
source-drain conductance (I$_{sd}$-V$_{g}$ curve) is shown in the inset of
Fig.{\nobreakspace}3. The left-most I-V curve in that inset corresponds to
full modulation doping after a deep soak in $\lambda $ = 1.77 $\mu $m light.

After the deep soak in $\lambda$ = 1.77 $\mu$m light to produce full channel
doping, the gate voltage is adjusted for a current around 0.6nAmp. The
device is then exposed to a photon flux at a wavelength of $\lambda$ = 1.3 $%
\mu$m (red curve labeled 1.3 $\mu$m in Fig.{\nobreakspace}3). The photon
exposure at $\lambda$ = 1.3 $\mu$m causes current to drop inexorably,
step-by-step, except for occasional upward spikes. Thus as a result of
trapped photoelectrons, the current is again pinched off, and the I$_{sd}$-V$%
_{g}$ curve was shifted toward positive gate voltages as shown in the
right-most curve of the inset in Fig.{\nobreakspace}3. At this pinch-off
condition, if the device was again exposed to $\lambda$ = 1.77 $\mu$m, the
channel current would be restored (blue curve labeled 1.7 $\mu$m in Fig.{%
\nobreakspace}3). The incident photon rate in the active window area for
both wavelengths is about 100{\nobreakspace}photons/s. Since the
absorptivity in the absorption layer is about 1\%, on average 1 photon/sec
is absorbed in the window area. Thus the quantum efficiency for producing
negative steps is estimated to be 1\%.

\begin{figure}
\includegraphics[width=8.1cm]{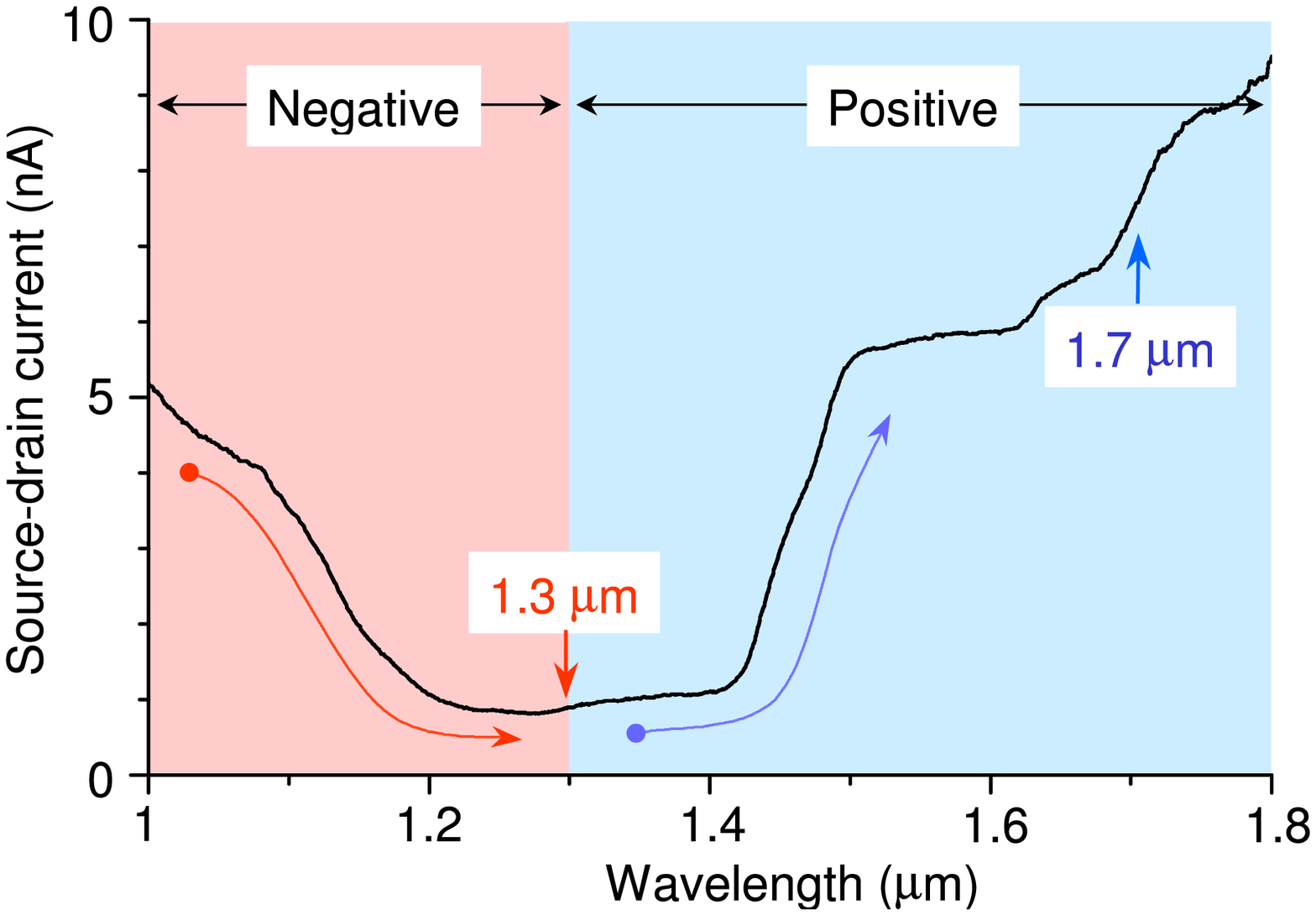}
\caption{\label{fig4} Spectral dependence of the
photoconductivity. The wavelength was swept from $\lambda$ = 1.0
$\mu$m to $\lambda$ = 1.8 $\mu$m while monitoring
the source-drain current. From $\lambda$ = 1.0 $\mu$m to $\lambda$ = 1.3 $%
\mu $m, the current monotonically decreases, which is the range of
negative photoconductivity. On the contrary, from $\lambda$ = 1.3
$\mu$m to $\lambda$ = 1.8 $\mu$m, the current monotonically
increases with increasing wavelength, which is the range of
positive photoconductivity. The cross-over point, $\lambda$ = 1.3
$\mu$m, corresponds to the bandgap in InGaAs quantum wells.}
\end{figure}

The current drop for $\lambda $ = 1.3 $\mu $m means that net negative charge
is trapped near the source/drain channel. The exposure to $\lambda $ = 1.77 $%
\mu $m photons is energetically able to cause only photo-ionization, because
the photon energy is smaller than any bandgaps. \newline
Detailed examination of the spectral dependence is not straightforward since
the channel conductance depends on the starting bias, and the full history
of spectral exposure. In Fig.{\nobreakspace}4 we start with an unpinched
channel, and sweep wavelength starting from $\lambda $ = 1 $\mu $m up to $%
\lambda $ = 1.8 $\mu $m over an 80second time period. First the current
monotonically decreases with increasing wavelength, corresponding to trapped
electrons, with no further decrease at around $\lambda $ = 1.3 $\mu $m, the
bandgap of the InGaAs quantum wells. Negative trapped charge at wavelengths
shorter than $\lambda $ = 1.3 $\mu $m is caused by photon absorption in the
absorption layer or the channel layer. The photoelectrons in the conducting
channel are mobile, and thus cannot contribute to trapped charge. Thus the
negative steps must originate from photoelectrons produced in the absorption
layer.

\begin{figure}
\includegraphics[width=8.1cm]{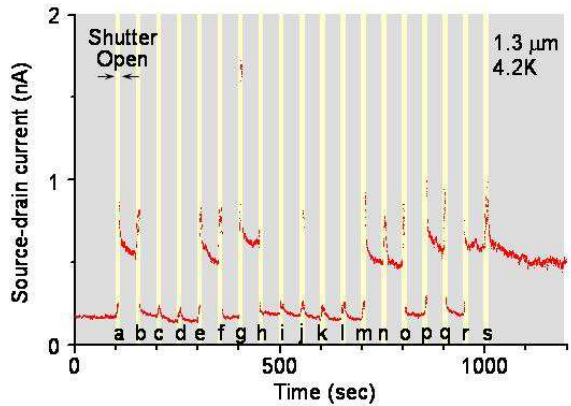}
\caption{\label{fig5} Bit wise current state switching near the
cross-over from positive to negative photoconductivity. The photon
source is gated to synchronize the
current steps with the photons. The shutter was repeatedly opened for $\sim$%
10sec every 50sec. The negative and positive photoconductivity
events (electron trapping and photo-neutralization), were balanced
by incomplete soaking at $\lambda$ = 1.77 $\mu$m. The current
alternates between a higher state and a lower state, the switching
induced by optical pulses. In the dark, the state was stable for
more than one hour. The photon number absorbed within the window
area is 30 photons in 10seconds, on average.}
\end{figure}

By having an incomplete initial soak in $\lambda $ = 1.77 $\mu $m radiation,
we can control the pinch off voltage in between -0.5V and +0.1V. Now, when
the pinch-off voltage is set to nearly zero, the $\lambda $ = 1.3 $\mu $m
photocurrent still shows steps but they are equally likely to be either
positive or negative. The incomplete photo-ionization of donors in the
initial state allows a balance between electron trapping and
photo-ionization. To make this phenomenon clear, we periodically opened the
optical shutter for 10 seconds in every 50 seconds, maintaining the SPT in a
balanced condition biased at $\sim $0 Volts. The resulting current pulses
are shown in Fig.{\nobreakspace}5. The optical shutter is open during the
time slots labeled a, b, c, etc., and closed during the intervening periods.
Successive optical pulses usually produced either electron trapping or
photo-ionization, alternating, depending on the previous state. Sometimes
multiple optical pulses were required before the state would alternate.
Within the 10sec optical pulse there might be a transient thermal response,
especially in time slot{\nobreakspace}g; but that returned to either of the
two alternating states after the optical pulse. The photon number absorbed
in the window area is{\nobreakspace}$\sim $30 on average within the 10sec
pulse. The estimated quantum efficiency is consistent with that in Fig.{%
\nobreakspace}3.

The switching behavior in Fig.{\nobreakspace}5 is due to photoelectron
trapping/de-trapping located either in; (a){\nobreakspace}the shallow
circular potential well between the window gates in the absorption layer, or
(b){\nobreakspace}at donor sites. In case{\nobreakspace}(b) the donors that
could contribute to trapping/de-trapping are the residual donors in the
absorption layer rather than those in the modulation doped layer. The
modulation-doped donors, which are located far below the channel, would only
produce a smooth increase in conductivity by photo-ionization as was seen in
Fig.{\nobreakspace}3 for $\lambda$ = 1.77 $\mu$m light. In either case{%
\nobreakspace}(a){\nobreakspace}or{\nobreakspace}(b), there are two possible
mechanisms for the positive steps in Fig.{\nobreakspace}5; photo-ionization
of the trapped electron, or annihilation of the trapped electron by injected
holes. The photo-ionization mechanism would require a specific
photo-ionization cross-section to be consistent with the rough equality
between trapping and de-trapping rates. On the other hand, annihilation by
photo-holes would require a hole trapping rate that is roughly coincident
with the electron trapping rate. Such an adjustment may have been made by
the adjustment of potential wells through the pinch-off voltage requirement
of Fig.{\nobreakspace}5.

In conclusion, we have trapped and safely stored single photoelectrons in a
window-gate double-quantum-well transistor structure. This Single
Photoelectron Transistor detector satisfies three key requirements for a
quantum repeater photo-detector. It has a wavelength suitable for optical
fibers, it safely traps and detects a single photoelectron, and the g$_{e}$%
-factors can be designed to satisfy the requirements for an entanglement
preserving photo-detector. The wavelength could be shifted to $\lambda$ =
1.55 $\mu$m, which is more preferable, by using strain engineered substrates
[15]. The Single Photoelectron Transistor announces the arrival of the
photoelectric charge, without disturbing the transfer of quantum information
from photon polarization to the electron spin state. We have yet to prove
the entanglement transfer, but we believe such a demonstration will be a
breakthrough for realizing long-distance quantum key distribution or long
distance teleportation.

The project is sponsored by the Defense Advanced Research Projects Agency \&
Army Research Office Nos. MDA972-99-1-0017 and DAAD19-00-1-0172. The content
of the information does not necessarily reflect the position or the policy
of the government, and no official endorsement should be inferred.



\begin{references}
\bibitem{Bennett92}  C.H. Bennett, F. Bessette, G. Brassard, L. Salvail, and
J. Smolin, J. Cryptology 5, 3 (1992).

\bibitem{Hiskett}  A.P. Hiskett, G. Bonfrate, G.S. Buller, and P.D.
Townsend, J. Modern Optics 48, 1957, (2001).

\bibitem{Bennett93}  C.H. Bennett, G. Brassard, C. Crepeau, R. Jozsa, A.
Peres, and W.K. Wootters, Phys. Rev. Lett. 70, 1895 (1993).

\bibitem{van Enk}  S.J. van Enk, J.I. Cirac and P. Zoller, Phys. Rev. Lett.
78, 4293 (1997).

\bibitem{Vrijen}  R. Vrijen and E. Yablonovitch, Physica E 10, 569 (2001).

\bibitem{shields}  A. J. Shields et al., Appl. Phys. Lett. 76, 3673 (2000).

\bibitem{kosaka02}  H. Kosaka et al., Phys. Rev. B 65, R201307 (2002).

\bibitem{rose}  A. Rose, Concepts in Photoconductivity and Allied Problems
(Krieger Publishing Co. Huntington New York, 1978).

\bibitem{blick98}  R.H. Blick, V. Gudmundsson, R.J. Haug, K. von Klitzing,
and K. Eberl, Phys. Rev. B 57, R12685 (1998).

\bibitem{blick95}  R.H. Blick, R.J. Haug, D.W. van der Weide, K. von
Klitzing, and K. Ebert, Appl. Phys. Lett. 67, 3924 (1995).

\bibitem{vaart}  N.C. van der Vaart et al., Phys. Rev. B 55, 9746 (1997).

\bibitem{komiyama}  S. Komiyama, O. Astafiev, V. Antonov, T. Kutsuwa, and H.
Hirai, Nature 403, 405 (2000).

\bibitem{awschalom}  D.D. Awschalom and J.M. Kikkawa, Physics Today 52, 33
(1999).

\bibitem{kosaka01}  H. Kosaka, A.A. Kiselev, F.A. Baron, K-W. Kim, and E.
Yablonovitch, Electron. Lett. 37, 464 (2001).

\bibitem{kiselev02}  A.A. Kiselev, K.W. Kim, and E. Yablonovitch, Appl.
Phys. Lett. 80, 2857 (2002).

\bibitem{chou}  W.Y. Chou, G.S. Chang, W.C. Hwang, and J.S. Hwang, J. Appl.
Phys. 83, 3690 (1998).
\end{references}
\end{document}